\documentclass[a4paper,12pt]{article}
\newcommand{\lapprox}{\stackrel{<}{\sim}}

\usepackage{epsfig}
\usepackage{graphicx}

\begin{document}
\baselineskip 20pt

\begin{huge}
\noindent
{\bf Enhanced paraconductivity-like fluctuations in the radio frequency spectra of ultracold Fermi atoms}
\end{huge}
\vspace{0.6cm}

\noindent
{\bf PIERBIAGIO PIERI*, ANDREA PERALI AND GIANCARLO CALVANESE STRINATI}
\vspace{0.3cm}

\noindent
{\bf Dipartimento di Fisica, Universit\`{a} di Camerino, I-62032 Camerino (MC), Italy}

\noindent
{\bf *e-mail: pierbiagio.pieri@unicam.it}
\vspace{1.5cm}

\noindent
{\bf Ultracold Fermi atoms allow the realization of the crossover from Bardeen-Cooper-Schrieffer (BCS) superconductivity to Bose- Einstein condensation (BEC)$^{1-3}$, by varying with continuity the attraction between fermions of different species$^{4}$.
In this context, radio frequency (RF) spectroscopy$^{5-7}$ provides a microscopic probe to infer the nature of fermionic pairing.
In the strongly-interacting regime, this pairing affects a wide temperature range comprising the critical temperature $T_{c}$, in analogy to the pseudogap physics for high-temperature superconductors.
By including what are known in superconductors as ``paraconductivity'' fluctuations, here we calculate the RF spectra above $T_{c}$ for fermions with balanced populations and compare them with available experimental data, thus revealing that the role of these fluctuations is considerably enhanced with respect to superconductors.
In addition, we extract from the spectra an energy scale associated with pairing and relate it to a universal quantity recently introduced for Fermi gases$^{8}$.}

\newpage 

It is widely appreciated at present that ultracold Fermi atoms represent ideal systems for deepening our understanding of condensed-matter systems, especially as far as the many-body quantum physics is concerned.
This is because the inter-particle interaction and species populations can be fine-controlled and tuned almost at will, in such a way that the system Hamiltonian is precisely known over a wide parameter range.
Since experimental techniques are able to provide diverse and accurate data, major theoretical efforts can be justified to attempt detailed comparison with experiments, particularly for those aspects that have remained unsettled in condensed matter.

In this respect, RF experiments$^{5-7}$ (especially in their tomographic version) provide fertile ground for exploring complementary aspects related to excitation gaps and pairing fluctuations, both below and above the critical temperature $T_{c}$ for the transition to the superfluid phase in these neutral systems.
A crucial aspect here is the occurrence of several hyperfine levels, which are split by a magnetic field and exhibit mutual Fano-Feshbach resonances that strongly amplify the two-body interaction when the field is varied$^{4}$.
The hyperfine levels act as components of a (hyper) spin and a spin flip is generated by the RF transition, so that a dynamical 
(two-particle) spin-correlation function is effectively measured by RF spectra$^{9}$.

Below $T_{c}$, theoretical interpretation of these spectra has revealed a competition between pairing-gap effects in the 
initial state of the transition and final-state effects, which tend to push the oscillator strength toward opposite sides of the 
spectrum$^{10}$.
In this case, the presence of a well-developed pairing gap at low temperature makes a BCS-RPA approach$^{11}$ 
for response functions sufficiently accurate to reproduce the experimental features$^{10}$.

When the temperature increases across $T_{c}$ the pairing gap gives the way to pairing fluctuations, and precursor 
effects of pairing appear in the normal phase above $T_{c}$.
These effects manifest themselves at the simplest level in the single-particle spectral function with a depression of the spectral weight (pseudogap) at low frequencies for small enough wave vectors
(the effect being amplified when approaching the BEC limit of the BCS-BEC crossover$^{12}$).
In high-temperature cuprate superconductors (where pairing is competing with other kinds of ordering), photoemission 
(ARPES) experiments have extensively been used to unravel the nature of the pseudogap in single-particle 
excitations$^{13}$.
Recently, attempts to disentangle single-particle excitations from RF (two-particle) spectra have also been 
made for ultracold Fermi gases$^{14}$, by resolving the wave vector of the photo-excited fermions.

Pairing fluctuations in the normal phase above $T_{c}$ are known to affect the two-particle response of superconductors.
In particular, for superconducting thin films the possibility of fluctuational creation of Cooper pairs has been shown 
to enhance their conductivity, an effect known as ``paraconductivity'' from the work of Aslamazov and Larkin$^{15}$ (AL). 
Another source of interaction of fermions with fluctuating Cooper pairs stems from the additional contribution to the
conductivity found by Maki$^{16}$ and Thompson$^{17}$ (MT).
The AL and MT contributions have been detected experimentally to produce small changes to the normal conductivity, 
which can be amplified by reducing the dimensionality$^{18}$.
While the AL contribution to the conductivity can be obtained theoretically by a time-dependent generalization of the 
Ginzburg-Landau equations$^{19}$, accounting for the MT contribution requires one to introduce finite-temperature 
Feynman diagrams for the current response function, a technique which yields also a third relevant contribution 
known as density-of-states (DOS) renormalization$^{20}$.

In this paper, we calculate the corresponding fluctuation contributions to the dynamical spin-correlation function associated with RF spectra, by using finite-temperature Feynman diagrams in the normal phase.
Our calculation reveals that, even in three dimensions, the role of pairing fluctuations can be drastically enhanced for the RF spin-correlation function of ultracold Fermi atoms with respect to the current-correlation function of superconductors.
The results of our calculation compare favorably with the available RF experimental data for Fermi gases with balanced populations under various coupling conditions.

The differences between the current- and RF spin-correlation functions originate from the different nature of the coupling to the external probe, the involvement of a third spin component and the wider coupling range that can be explored via Fano-Feshbach resonances with Fermi atoms.
Apart from these differences (which yet make the MT contribution vanish identically for the RF spin-correlation function), the topological structure of the Feynman diagrams for the AL and DOS contributions remains the same for both calculations
(see the Methods section).

For the RF spectra of interest here, two spin components ($\alpha$ and $\beta$) are initially equally populated while a third ($\gamma$) is empty.
For a transition which flips $\beta$ into $\gamma$, the relevant scattering lengths are then $a_{\alpha \beta}$ for the initial 
($i$) and $a_{\alpha \gamma}$ for the final ($f$) state.
In our calculation, the DOS contribution contains only pairing fluctuations related to $a_{\alpha \beta}$ (initial-state effects), while the AL contribution contains in addition pairing fluctuations related to $a_{\alpha \gamma}$ (final-state effects) - see the Methods section. For this reason, we refer to $a_{\alpha \beta}$ as $a_{i}$ and to $a_{\alpha \gamma}$ as $a_{f}$.
Experiments have been able to explore different combinations of these spin components, thereby enlarging the accessible coupling ranges.
Figure 1 reports the corresponding locations of the possible experimental coupling values for $^6$Li in the 
$((k_{F}a_{i})^{-1},(k_{F}a_{f})^{-1})$ plane, where $k_{F}$ is the Fermi wave vector.

Comparisons between the results of our numerical calculations and the experimental spectra for $^6$Li are reported in 
Fig. 2, resulting in quite good agreement between theory and experiments.
Recall that, in the absence of pairing fluctuations, a single delta-spike at zero frequency would absorb the whole spectra 
weight (whose value is dictated by a sum rule for the spectral area$^{10}$ and is not changed by further inclusion of 
$a_{i}$ and $a_{f}$).
In all panels, the left peak is associated with the presence of a localized (bound) final state, which typically absorbs a spectral weight comparable to the right peak.
The latter originates from continuum transitions and comprises the high-frequency tail. 
For these reasons, the final-state effects associated with the AL contribution play an essential role in all cases
(see also Supplementary Fig. 1).
Nonetheless, when the bound and continuum peaks are sufficiently apart, the continuum part of the spectra \emph{appears} 
to be only slightly modified by the inclusion of final-state effects \emph{provided} the peak heights (obtained by DOS only and by DOS plus AL) are brought to coincide.
These considerations justify \emph{a posteriori} the procedure adopted in ref. 7 to fit the continuum part of the experimental spectra disregarding final-state effects.

Experimentally, the motivation for exploring RF spectroscopy with ultracold Fermi atoms$^{5}$ stemmed from the expectation 
of extracting the value of a pairing gap directly from the (continuum part of the) spectra. 
This would be possible within a BCS description al low temperature, since in this case the continuum edge occurs at 
$\Delta^{2}/(2E_{F})$ where $\Delta$ is the BCS gap parameter and $E_{F} = k_{F}^{2}/(2 m)$ the Fermi energy ($m$ being 
the fermion mass).
Such a simplified picture, however, does not apply to the unitarity regime about $(k_{F} a_{i})^{-1}=0$ owing to strong-coupling effects in the initial and/or final states of the transition.
Accordingly, no procedure has been suggested so far for extracting from the RF spectra an energy scale associated with pairing interaction in systems with balanced populations$^{23}$.

The present analysis allows us to identify such an energy scale when the bound and continuum peaks of an RF spectrum are 
sufficiently far apart (corresponding to the green area of Fig. 1).
In this case, there exists a window on the high-frequency side of the continuum peak where the RF signal behaves like
$B \, (\omega/E_{F})^{-3/2}$ with $B = A \, (3/2^{5/2}) \, (\Delta_{\infty}/E_{F})^{2}$.
Here, $A$ stands for the area of the continuum peak once the total area of the RF spectrum is taken to be unity.
Typically, this frequency window extends between a few times $E_{F}$ and $(m a_{f}^{2})^{-1}$, past which the spectral 
tail is dominated by final-state effects and decays like $\omega^{-5/2}$.
The above expression is written in analogy with the high-frequency behavior of the RF signal within the BCS approximation, 
in which $\Delta$ appears in place of $\Delta_{\infty}$.
The energy scale $\Delta_{\infty}$ can be further related to the asymptotic behavior $C / \mathbf{k}^{4}$ of the 
wave-vector distribution function $n(\mathbf{k})$, where $C$ is the ``contact intensity'' introduced in ref. 8 which  
enters several quantities of a Fermi gas in a universal way$^{24,25}$. 
From our theory we identify $C = ( m \, \Delta_{\infty})^{2}$ (see the Methods section).

In Fig. 3a the theoretical RF spectra show a plateau whose height is $(3/2^{5/2}) \, (\Delta_{\infty}/E_{F})^{2}$ according 
to the above argument.
This height is compared with the horizontal line, obtained by the value of $\Delta_{\infty}/E_{F}$ 
resulting from our independent calculation of the contact intensity $C$ through $n(\mathbf{k})$.
In Fig. 3b we apply the same procedure to the experimental RF spectra.
Although the data are somewhat scattered in the frequency window of interest, their tendency to approach a plateau 
results from the figure and allows us to obtain a numerical estimate for $\Delta_{\infty}$.
This procedure could be implemented in future experiments to map the value of $\Delta_{\infty}$ in 
extended temperature and coupling ranges.
 
\newpage

\noindent
{\bf METHODS}
\vspace{0.3cm}

\begin{footnotesize}
\noindent
{\bf RF SPECTRUM  AND THE SPIN CORRELATION FUNCTION}
\end{footnotesize}

\noindent
 The RF spectrum is obtained theoretically within linear-response theory as:

\[
\delta \langle I(\omega_{\mathrm{th}}) \rangle \, = \, - \, 2 \, g^{2} \int \! d\mathbf{r} \,
d\mathbf{r'} \, \, \mathrm{Im} \{\Pi_{\beta \gamma}^{R}(\mathbf{r},\mathbf{r'};\omega_{\mathrm{th}})\} 
\]

\noindent
where $g$ is the coupling constant of the atomic ($\beta \rightarrow \gamma$) transition,
$\Pi_{\beta \gamma}^{R}(\mathbf{r},\mathbf{r'};\omega_{\mathrm{th}})$ is the Fourier transform of the retarded
correlation function $\Pi_{\beta \gamma}^{R}(\mathbf{r},\mathbf{r'};t - t') = - i \theta(t - t') 
\langle [ B_{\beta \gamma}(\mathbf{r},t), B^{\dagger}_{\beta \gamma}(\mathbf{r'},t') ]  \rangle$ taken at the shifted frequency 
$\omega_{\mathrm{th}} = \omega_{RF} + \mu_{\beta} - \mu_{\gamma}$ with respect to the RF frequency $\omega_{RF}$
(we set $\hbar = 1$ throughout).
Here, $B_{\beta \gamma}^{\dagger}(\mathbf{r}) = \psi_{\gamma}^{\dagger}(\mathbf{r}) \psi_{\beta}(\mathbf{r})$ is the transition operator in terms of the field operators $\psi_{\beta/\gamma}$ at the spatial point $\mathbf{r}$, $\mu_{\beta}$ and 
$\mu_{\gamma}$ are the chemical potentials for the initial (equally populated) and final (empty) levels, in the order, and $\langle \cdots \rangle$ stands for a thermal average.

Following a standard procedure$^{26}$, the retarded correlation function is conveniently calculated via its Matsubara counterpart

\[
\Pi_{\beta \gamma}(\mathbf{r},\mathbf{r'};\Omega_{\nu}) \, = \, \int_{0}^{\beta} \! 
 d\tau \, e^{i \Omega_{\nu} \tau}
\langle \mathrm{T}_{\tau} [ \psi_{\beta}(\mathbf{r'},0) \psi_{\beta}^{\dagger}(\mathbf{r},\tau^{+}) 
\psi_{\gamma}(\mathbf{r},\tau) \psi_{\gamma}^{\dagger}(\mathbf{r'},0^{+})  ] \rangle \,\, ,
\]

\noindent
for this expression admits a representation in terms of Feynman diagrams.
Here, $\mathrm{T}_{\tau}$ is the time-ordering operator for imaginary time $\tau$ and $\Omega_{\nu} = 2 \pi \nu T$ ($\nu$ integer) is a bosonic Matsubara frequency at temperature $T$ (we set $k_{B} = 1$ throughout).
Above $T_{c}$, the relevant Feynman diagrams for the two-particle response are the DOS and AL contributions depicted 
in Fig. 4, which result once the single-particle self-energy with pairing fluctuations is adopted.
Analytic continuation from $i \Omega_{\nu}$ to $\omega_{\mathrm{th}} + i \eta$ ($\eta$ being a positive infinitesimal) is eventually required.
To compare with the experimental spectra, we have to convert from $\omega_{\mathrm{th}}$ to 
 $\omega = \omega_{RF} - \omega_{\gamma \beta}$ where $\omega_{\gamma \beta}$ is the Bohr frequency 
 of the atomic transition.
Besides being appropriate to the BCS ($(a_{\alpha \beta} k_{F})^{-1} \lapprox -1$) and unitarity
($(a_{\alpha \beta} k_{F})^{-1} \approx 0$) regimes, the DOS and AL contributions are also able to recover the two-body (molecular) calculation of ref. 27 when the BEC regime ($1 \lapprox (a_{\alpha \beta} k_{F})^{-1}$) is approached.
(In the text, we have set $a_{\alpha \beta} = a_{i}$.)
\vspace{0.5cm}

\begin{footnotesize}
\noindent
{\bf NUMERICAL PROCEDURES}
\end{footnotesize}

\noindent
Due to the involved structure of the AL contribution (see Fig. 4b), for the external Matsubara frequency $\Omega_{\nu}$ analytic continuation from the points $i \Omega_{\nu}$ on the imaginary axis to (just above) the real frequency axis cannot be achieved by the elementary methods that proved sufficient$^{10}$ for the BCS-RPA approach below $T_{c}$.
The delicate analytic continuation at finite temperature which is required for the AL diagram has been performed in the literature with the resort to approximations$^{20}$ which do not apply to the present case, when experimental RF spectra with nontrivial frequency structures need be accounted for in a variety of situations.

To this end, we have performed the analytic continuation of the AL contribution resorting to Pad\'{e} approximants$^{28}$, whereby the dependence on the external frequency $\Omega_{\nu}$ is approximated by the ratio of two polynomials (the denominator having one additional power with respect to the power $M$ of the numerator).
The $2 M$ unknown coefficients have been determined by sampling over about 450 Matsubara frequencies (not necessarily equally spaced).
The reliability of the numerical results has been tested against the DOS contribution above $T_{c}$ as well as 
the BCS-RPA calculation below $T_{c}$, which can both be obtained by direct analytic continuation without resorting to 
Pad\'{e} approximants.
Additional independent tests on the numerical procedure are the values of the sum rules for the spectral area$^{10}$ 
and its first moment$^{9}$, which are reproduced within $1 \%$ and $10 \%$, in the order.

This procedure of analytic continuation was avoided in ref. 29 while addressing the calculation 
of RF spectra above $T_{c}$, by effectively extending the BCS-RPA calculation to temperatures between the actual $T_{c}$ and the corresponding temperature obtained at the mean-field level (often referred to as the temperature $T^{*}$ at which pairing exhausts its effects). 
\vspace{0.5cm}

\begin{footnotesize}
\noindent
{\bf THE ENERGY SCALE $\Delta_{\infty}$ OF THE PAIRING INTERACTION}
\end{footnotesize}

\noindent
The most direct way to introduce the quantity $\Delta_{\infty}$ is through the coefficient $(3/2^{5/2}) \, \Delta_{\infty}^{2}$ 
of the $\omega^{- 3/2}$ behavior of the DOS and AL contributions for large $\omega$, which identifies

\[
\Delta_{\infty}^{2} = \int \! \frac{d \mathbf{q}}{(2 \pi)^{3}} \, \frac{1}{\beta} \, \sum_{\Omega_{\nu}} \,
e^{i \Omega_{\nu} \eta} \,\, \Gamma_{\alpha \beta}(\mathbf{q},\Omega_{\nu}) 
\]

\noindent
where $\mathbf{q}$ is a wave vector and $\Gamma_{\alpha \beta}$ is the pairing propagator of Fig. 4c.
This expression is conveniently evaluated by using the spectral representation$^{30}$ of $\Gamma_{\alpha \beta}$.
In the BCS regime $\Delta_{\infty}$ coincides with the absolute value $2 \pi |a_{i}| n/m$ of the mean-field shift for 
temperatures smaller than $(m a_{i}^{2})^{-1}$, while in the BEC regime the relation $\Delta_{\infty}^{2} = 4 \pi n/(m^{2} a_{i})$
holds in analogy with the BCS mean-field result.
Close to $T_c$ at unitarity, the numerical value $0.75 E_{F}$ of $\Delta_{\infty}$ is comparable with the value $0.8 E_{F}$ of the pseudogap extracted from the single-particle spectral function$^{12}$.
The temperature dependence of $\Delta_{\infty}$ is rather weak for all coupling regimes, extending up to several times the 
Fermi temperature $T_{F}$.

The energy scale $\Delta_{\infty}$ is also related to the asymptotic behavior of the wave-vector distribution function
$n(\mathbf{k})$ as obtained from the fermionic single-particle propagator $\beta$-$\beta$ dressed by the self-energy 
$\Sigma_{\beta}$ of Fig. 4c.
We get $n(\mathbf{k}) = (m \, \Delta_{\infty})^{2} /\mathbf{k}^{4}$ where $\Delta_{\infty}^{2}$ is defined as above.
From this expression we identify $( m \, \Delta_{\infty})^{2}$ with the contact intensity $C$ of ref. 8.

\newpage

\noindent
{\bf References}
                                                                                                                                                                                                                                                                                                                                                                                                        
\noindent
1. Eagles, D. M. Possible pairing without superconductivity at low carrier
    concentrations in bulk and thin-film superconducting semiconductors. 
    \emph{Phys. Rev.} {\bf186,} 456-463 (1969).
    
\noindent    
2. Leggett, A. J. in \emph{Modern Trends in the Theory of Condensed Matter}
    (\emph{Proc. 16th Karpacz Winter School Theor. Phys.}) 
    (eds Pekalski, A. $\&$ Przystawa, J.) 13-27 (Springer, 1980).
 
\noindent    
3. Nozi\'{e}res, P. $\&$ Schmitt-Rink, S. 
    Bose condensation in an attractive fermion gas: from weak to strong coupling 
    superconductivity. 
    \emph{J. Low Temp. Phys.} {\bf 59,} 195-211 (1985).
    
\noindent
4. See, e.g., Ketterle, W. $\&$ Zwierlein, M. W. in \emph{Ultra-Cold Fermi Gases}  
    (\emph{Proc. Internat. School Phys. ÔEnrico FermiÕ, Course 164}) 
    (eds Inguscio, M., Ketterle, W. $\&$ Salomon, C.) 95-287 (IOS Press, 2008).    

\noindent
5. Chin, C. \emph{et al.} Observation of the pairing gap in a strongly interacting Fermi 
    gas. \emph{Science} {\bf 305,} 1128-1130 (2004).
    
\noindent
6. Shin, Y., Schunck, C. H., Schirotzek, A. $\&$ Ketterle, W. 
    Tomographic rf spectroscopy of a trapped Fermi gas at unitarity. 
    \emph{Phys. Rev. Lett.} {\bf 99,} 090403 (2007).

\noindent    
7. Schunck, C. H., Shin, Y., Schirotzek, A. $\&$ Ketterle, W.  
    Determination of the fermion pair size in a resonantly interacting superfluid.
    \emph{Nature} {\bf 454,} 739-743 (2008). 
    
\noindent    
8. Tan, S. Large momentum part of a strongly correlated Fermi gas.  
    \emph{Ann. Phys.} doi:10.1016/j.aop.2008.03.005 (2008).

\noindent    
9. Yu, Z. $\&$ Baym, G. Spin-correlation functions in ultracold paired atomic-fermion systems:
    Sum rules, self-consistent approximations, and mean fields.
    \emph{Phys. Rev. A} {\bf 73,} 063601 (2006).
    
\noindent    
10. Perali, A., Pieri, P. $\&$ Strinati, G. C. Competition between final-state and pairing-gap 
      effects in the radio-frequency spectra of ultracold Fermi atoms.
      \emph{Phys. Rev. Lett.} {\bf 100,} 010402 (2008). 
      
\noindent    
11. Schrieffer, J. R. \emph{Theory of Superconductivity} Ch. 8 (W. A. Benjamin, New York, 1964).  

\noindent
12. Perali, A., Pieri, A., Strinati, G. C. $\&$ Castellani, C. 
      Pseudogap and spectral function from superconducting fluctuations to the bosonic limit. 
      \emph{Phys. Rev. B} {\bf 66,} 024510 (2002).

\noindent
13. Damascelli, A. Probing the electronic structure of complex systems by ARPES.
      \emph{Phys. Scr.} {\bf T109,} 61-74 (2004).

\noindent
14. Stewart, J. T., Gaebler, J. P. $\&$ Jin, D. S.
      Using photoemission spectroscopy to probe a strongly interacting Fermi gas.
       \emph{Nature} {\bf 454,} 744-747 (2008).
       
\noindent
15. Aslamazov, L. G. $\&$ Larkin, A. I.
     Effect of fluctuations on the properties of a superconductor above the critical temperature.
     \emph{Soviet Solid State} {\bf 10,} 875-880 (1968).
     
\noindent
16. Maki, K.
      Critical fluctuation of the order parameter in a superconductor.
      \emph{Prog. Theor. Phys. (Kyoto)} {\bf 40,} 193-200 (1968).
      
\noindent
17. Thompson, R. S.
      Microwave, flux flow, and fluctuation resistance of dirty type-II superconductors.
      \emph{Phys. Rev. B} {\bf 1,} 327-333 (1970).                  

\noindent
18. Skocpol, W. J. $\&$ Tinkham, M. Fluctuations near superconducting phase transitions.
      \emph{Rep. Prog. Phys.} {\bf 38,} 1049-1097 (1975).

\noindent    
19. Abrikosov, A. A. \emph{Fundamentals of the Theory of Metals} Ch. 19
      (North-Holland, Amsterdam, 1988).    
      
\noindent
20. Varlamov, A. $\&$ Larkin, A. \emph{Theory of Fluctuations in Superconductors} Ch. 7 
      (Oxford Scholarship Online, Oxford, 2007).
      
\noindent
21. Bartenstein, M.  \emph{et al.} 
      Precise determination of $^6$Li cold collision parameters by radio-frequency spectroscopy on weakly bound molecules.
      \emph{Phys. Rev. Lett.} {\bf 94,} 103201 (2005).     
      
\noindent
22. Simonucci, S., Pieri, P. $\&$ Strinati, G. C.
      Broad vs. narrow Fano-Feshbach resonances in the BCS-BEC crossover with trapped Fermi atoms.
      \emph{Europhys. Lett.} {\bf 69,} 713-718 (2005).
      
\noindent
23. Schirotzek, A., Shin, Y., Schunck, C. H. $\&$ Ketterle, W.   
      Determination of the superfluid gap in atomic Fermi gases by quasiparticle spectroscopy.  
      Preprint at (http://arxiv.org/abs/0808.0026v2)(2008).
      
\noindent      
24. Braaten, E. $\&$ Platter L.
      Exact relations for a strongly interacting Fermi gas from the operator product expansion.
      \emph{Phys. Rev. Lett.} {\bf 100,} 205301 (2008).

\noindent      
25. Werner, F., Tarruell L. $\&$ Castin, Y.
      Number of closed-channel molecules in the BEC-BCS crossover.
      Preprint at (http://arxiv.org/abs/0807.0078v1)(2008).
      
\noindent
26. Fetter, A. L. $\&$ Walecka, J. D. \emph{Quantum Theory of Many-Particle Systems} Ch. 9
     (McGraw-Hill, New York, 1971).
      
\noindent      
27. Chin, C. $\&$ Julienne P. S.
      Radio-frequency transitions on weakly bound ultracold molecules.
      \emph{Phys. Rev. A} {\bf 71,} 012713 (2005). 

\noindent
28. Vidberg, H. J. $\&$ Serene, J. W.
      Solving the Eliashberg equations by means of $N-$point Pad\'{e} approximants.
      \emph{J. of Low. Temp. Phys.} {\bf 29,} 179-192 (1977).
            
\noindent      
29. He, Y., Chien, C. C., Chen, Q. $\&$ Levin K. 
      Temperature and final state effects in radio frequency spectroscopy experiments on atomic Fermi
      gases.   
      Preprint at (http://arxiv.org/abs/08041429v1)(2008). 
      
\noindent
30. Pieri, P., Pisani, L. $\&$ Strinati, G. C.
      BCS-BEC crossover at finite temperature in the broken-symmetry phase.  
      \emph{Phys. Rev. B} {\bf 70,} 094508 (2004).
      
\vspace{0.5cm}

\noindent
{\bf Acknowledgements}

\noindent
This work was partially supported by the Italian MUR under contract PRIN-2007 
``Ultracold Atoms and Novel Quantum Phases''.
This paper is dedicated to the memory of Professor Franco Bassani.      


\newpage
\begin{figure}[htbp]
\centering
\includegraphics[scale=0.8,angle=0]{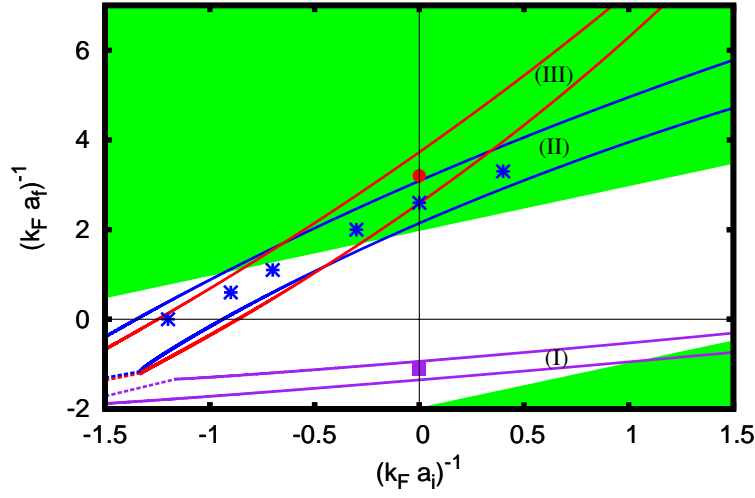}
\caption{\textbf{Allowed coupling values for} $\mathbf{^{6}Li}$ \textbf{in the} $(\mathbf{(k_{F}a_{i})^{-1},(k_{F}a_{f})^{-1}})$ \textbf{plane.}
RF experiments realized with ultracold $^{6}$Li atoms correspond to the three following combinations of scattering lengths: 
(I) for $(a_{i}=a_{12},a_{f}=a_{13})$, (II) for $(a_{i}=a_{13},a_{f}=a_{12})$, and (III) for $(a_{i}=a_{13},a_{f}=a_{23})$.
Using the values of these scattering lengths vs. the magnetic field given in ref. 21 and the experimentally accessible values
of the Fermi wave vector $k_{F} = (3 \pi^{2} n)^{1/3}$ where $n$ is the (total) density at the trap center (typically,
$k_{F} \approx 2.7 \times 10^{-4} a_{0}^{-1}$ where $a_{0}$ is Bohr radius), we construct the three stripes labeled (I), (II), 
and (III) inside which RF spectra can be collected, whose widths correspond to an estimated $40\%$ variation on the values 
of $k_{F}$.
The symbols correspond to the three different sets of tomographic RF spectra reported in ref. 7.
The green area indicates the region $2 \lapprox |(k_{F}a_{i})^{-1} - (k_{F}a_{f})^{-1}|$ where it is possible to extract the 
quantity $\Delta_{\infty}$ (see below) from the high-frequency tail of the RF spectra.}
\end{figure}

\newpage
\begin{figure}[htbp]
\centering
\vspace{-2.6truecm}
\includegraphics[scale=0.4,angle=0]{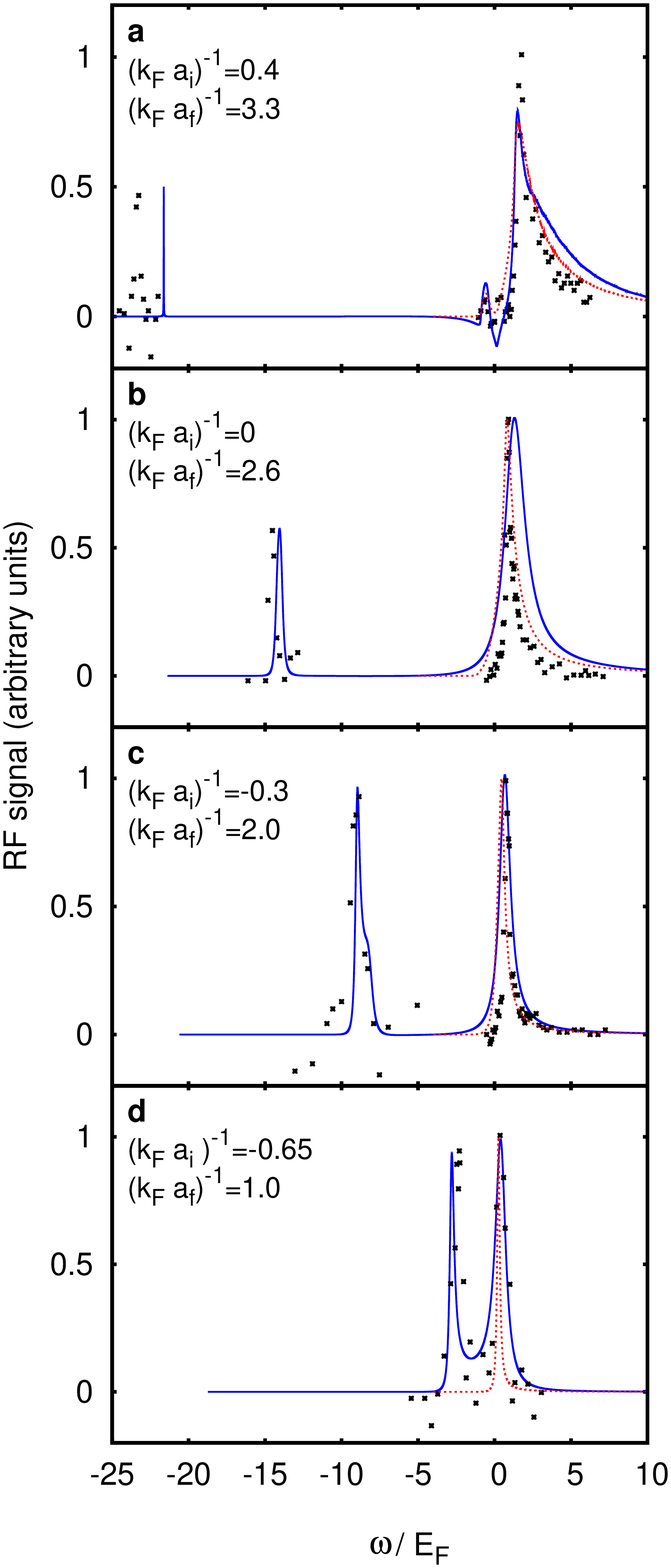}
\vspace{-.4truecm}
\caption{\small{\textbf{Comparison between theoretical and experimental RF spectra.}
The experimental spectra from Figs. 4a-d of ref. 7 (black squares) are compared with our theoretical calculations, obtained
with the inclusion of pairing fluctuations due to $a_{i}$ (DOS contribution - red dotted curves) and with the additional inclusion of pairing fluctuations due to $a_{f}$ (DOS plus AL contributions - blue full curves).
The values $a_{i}$, $a_{f}$, and $k_{F}$ are taken from the experiments.
The experimental spectra (which are tomographic for panels \textbf{a-c} and trap-averaged for panel \textbf{d}) span approximately the temperature range $(0.7 \, T_{c} - T_{c})$, and reveal a smooth evolution across $T_{c}$.
In all cases, the theoretical spectra are calculated for a homogeneous system in the normal phase at a temperature 
$T \, (=1.1 T_{c})$ slightly above $T_{c}$. 
In panels \textbf{a-c}, the bound (left) and continuum (right) peaks of the experimental spectra were independently normalized 
to the corresponding peak heights, while in panel \textbf{d} a single normalization was used$^{7}$.
We have adopted the same procedure for the comparison.
In addition, our calculation provides for the relative weight of the bound peak the values $59\%$ (panel \textbf{a}), $44\%$ (panel \textbf{b}), and $33\%$ (panel \textbf{c}).
In panel \textbf{a} the discrepancy between the positions of the experimental and theoretical bound peaks can be accounted 
for by a molecular multi-channel calculation$^{22}$.}}  
\end{figure}

\newpage
\begin{figure}[htbp]
\centering
\includegraphics[scale=0.6,angle=0]{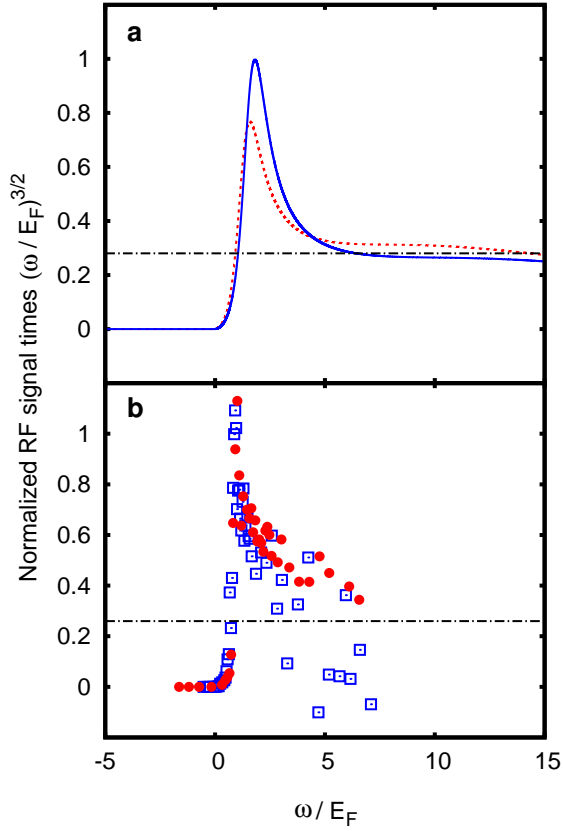}
\caption{\textbf{Procedure for extracting} $\mathbf{\Delta_{\infty}}$ \textbf{from the tail of the RF spectra.}
\textbf{a}, Theoretical RF spectra normalized to the weight $A$ of the continuum peak and multiplied by 
$(\omega/E_F)^{3/2}$.
The blue full curve is for $(k_{F} a_{f})^{-1} = 2.6$ (with $A = 56\%$) and the red dotted curve for $(k_{F} a_{f})^{-1}= 3.2$ 
(with $A = 63\%$), while $(k_{F} a_{i})^{-1}=0$ and $T = 1.1 T_{c}$ in both cases.
The horizontal line corresponds to the value $\Delta_{\infty} = 0.73 E_{F}$ obtained at the same temperature.
\textbf{b}, An analogous procedure is applied to the corresponding experimental spectra, by normalizing the area of the
continuum peak to unity.
Blue empty squares are from Fig. 2b of ref. 7 with $(k_{F} a_{f})^{-1} = 2.6$ and red full dots are from Supplementary Fig. 1 
of ref. 7 with $(k_{F} a_{f})^{-1} = 3.2$, while $(k_{F} a_{i})^{-1}=0$ and the temperature is below (but close to) $T_{c}$ in 
both cases. 
The data yield approximately the value $0.25 \pm 0.10$ for the height of the plateau, from which we get
$\Delta_{\infty}/E_{F} = 0.69^{+0.12}_{-0.16}$ with no input from theory.}  
\end{figure}

\newpage
\begin{figure}[htbp]
\centering
\includegraphics[scale=0.50,angle=0]{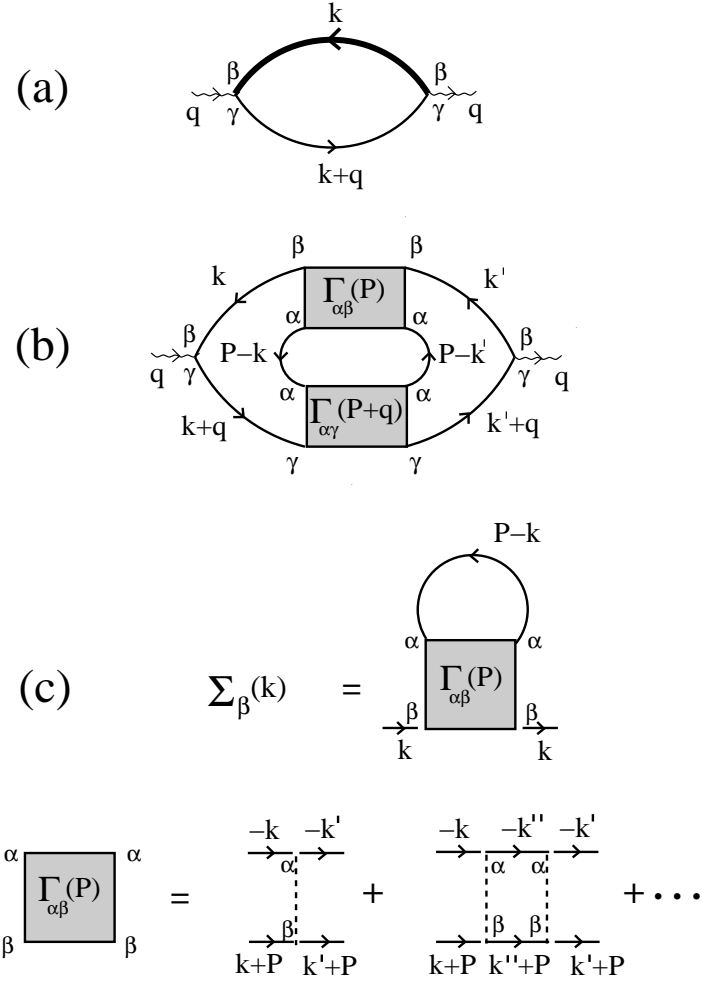}
\caption{\textbf{Diagrammatic representation of the spin correlation function above} $\mathbf{T_{c}}.$
\textbf{a}, DOS contribution where the fermionic $\beta$-$\beta$ single-particle propagator (upper line) is dressed
by the self-energy $\Sigma_{\beta}$, which includes pairing fluctuations of the initial state $\beta$ with its mate $\alpha$ 
through the pairing propagator $\Gamma_{\alpha \beta}$.
\textbf{b}, AL contribution which includes, in addition, pairing fluctuations of the final state $\gamma$ with the mate $\alpha$
left behind.
\textbf{c}, Self-energy $\Sigma_{\beta}$ and pairing propagator $\Gamma_{\alpha \beta}$ between spins $\alpha$ 
and $\beta$ in the initial state ($\Gamma_{\alpha \gamma}$ between spins $\alpha$ and $\gamma$ in the final state 
is similarly obtained).
Solid and dashed lines stand for fermionic single-particle propagators and interactions, respectively, while labels attached 
to the end points identify the relevant spins. 
All diagrams are drawn in four-momentum space.
Only the single-particle propagator $\beta$-$\beta$ in panel \textbf{a} is dressed by self-energy insertions.}  
\end{figure}

\newpage
\begin{figure}[htbp]
\centering
\includegraphics[scale=0.7,angle=0]{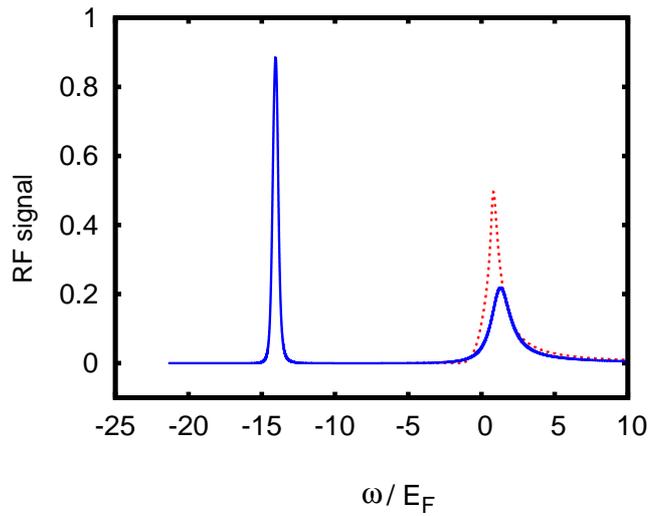}
\vspace{1truecm}
\caption{\textbf{Supplementary Figure 1 Comparison between the DOS only and DOS plus AL contributions to the RF spectra on an absolute scale.}
The DOS only (red dotted curve) and the DOS plus AL (blue full curve) contributions to the RF spectra are compared with
each other on an absolute scale, defined by the total area of each curve being unity.
The values of the couplings and temperature correspond to Fig. 2b of the text, where the two curves were instead normalized
so that the heights of the corresponding continuum peaks coincide with each other.
The present absolute comparison shows an even more marked difference between the two cases, when final-state effects 
are either neglected (DOS only) or included (DOS plus AL).}  
\end{figure}

\end{document}